\documentclass[a4paper,twoside,twocolumn,english,showpacs]{revtex4}
\usepackage{times}
\usepackage[OT1]{fontenc}
\usepackage[latin1]{inputenc}
\usepackage{amsmath}
\usepackage{graphicx}
\usepackage{amssymb}

\makeatletter

\usepackage{babel}
\makeatother
\begin{document}

\title{The Projected Gross-Pitaevskii Equation for harmonically confined
Bose gases}

\author{P. Blair Blakie$^{1}$ and Matthew J. Davis$^{2}$}

\affiliation{$^{1}$University of Otago, P.O. Box 56, Dunedin, New Zealand\\
\noindent$^{2}$ARC Centre of Excellence for Quantum-Atom Optics,
School of Physical Sciences, University of Queensland, Brisbane, QLD
4072, Australia}

\date{20th October 2004.}

\pacs{03.75.-b, 03.75.Hh}

\begin{abstract}
We extend the Projected Gross Pitaevskii equation formalism of Davis
\emph{et al.} {[}Phys. Rev. Lett. \textbf{87}, 160402 (2001){]} to
the experimentally relevant case of harmonic potentials. We outline
a robust and accurate numerical scheme that can efficiently simulate
this system. We apply this method to investigate the equilibrium properties
of a harmonically trapped three-dimensional Bose gas at finite temperature,
and consider the dependence of condensate fraction, position and momentum
distributions, and density fluctuations on temperature. We apply the
scheme to simulate an evaporative cooling process in which the preferential
removal of high energy particles leads to the growth of a Bose-Einstein
condensate. \textbf{}We show that a condensate fraction can be inferred
during the dynamics even in this non-equilibrium situation.
\end{abstract}
\maketitle

\section{Introduction}

The near zero temperature formalism for Bose-Einstein condensates
is based upon the well-established Gross-Pitaevskii equation (GPE)
that describes the macroscopically occupied condensate orbital. It
has been proposed that the GPE can also be used to model the non-equilibrium
dynamics of finite temperature Bose gases \cite{Svistunov1991,Kagan1992,Kagan1994,Kagan1997}.
The essential idea is that highly occupied modes of a quantum field
are well described by a classical field, as is well-known for the
case of electromagnetic fields. Recently, calculations have been performed
by several groups using classical fields \cite{Marshall1999a,Davis2001a,Davis2002a,DavisTemp,Goral2001a,Goral2002a,Schmidt2003a,Brewczyk2004a}
and have shown the usefulness of this approach. One of the key advantages
of the classical field method is that it is non-perturbative; however
care must be taken to appropriately cutoff the spectrum to avoid an
ultraviolet catastrophe, analogous to that occurring in the Rayleigh-Jeans
theory of blackbody radiation. Ideally this cutoff can be made from
\emph{a priori} thermodynamic analysis of the system. In the finite
temperature formalism of Davis \emph{et al.} \cite{Davis2001b,Davis2001a,Davis2002a},
the cutoff is explicitly incorporated through the use of a projector
that is diagonal in the single particle basis of the system Hamiltonian.
This ensures that a consistent energy cutoff is established, and provides
a natural separation of the system into regions of low energy modes
(the highly occupied \emph{coherent region}) and high energy modes
(the sparsely occupied \emph{incoherent region}). Ignoring the incoherent
region and its coupling to the coherent region, the equation of motion
for the low energy modes is termed the Projected Gross-Pitaevskii
Equation (PGPE). By itself the PGPE can only provide a partial description
of the system, however it contains the modes that are most \textbf{}significantly
modified by the effects of interactions and are difficult to include
quantitatively in traditional kinetic theories. Thus the PGPE by itself
can be a useful tool for providing insight into the evolution of ultra-cold
Bose gases.

The Stochastic GPE formalism developed by Gardiner and coworkers \cite{Gardiner2002a,Gardiner2003a}
proposes a practical scheme for coupling the coherent and incoherent
regions in a manner suitable for non-equilibrium calculations. To
account for the interaction with the incoherent region, this formalism
introduces noise terms into the PGPE that transfer energy and particles
into and out of the coherent region. 

A related technique originating in quantum optics is the phase space
method known as the truncated Wigner approximation. This was first
applied to ultra-cold Bose gases by Steel \emph{et al.} \cite{Steel1998a},
and further developed by Sinatra and co-workers \cite{Sinatra2000a,Sinatra2001a,Sinatra2002a}.
A number of recent calculations using this method can be found in
Refs. \cite{Polkovnikov2003a,Polkovnikov2004a,Lobo2004a,Norrie2004a,Isella2004a}.

The main purpose of this paper is to develop the PGPE approach for
the experimentally relevant case of three dimensional harmonic traps.
To do this we have adapted a recent numerical scheme by Dion and Canc\`es
\cite{Dion2003a} by introducing an explicit energy cutoff in the
single particle (harmonic oscillator) basis. We use the scheme to
examine the properties of trapped gases as a function of the temperature,
such as condensate fraction and density fluctuations. While some of
these properties were studied in Ref. \cite{Goral2002a}, we believe
that our numerical method is superior and our results are free from
the significant uncertainties found in this paper. Finally, as a non-equilibrium
application we use the PGPE to simulate the formation of a Bose-Einstein
condensate in an evaporatively cooled thermal cloud.

\section{Formalism \label{sec:Formalism}}

The theoretical formalism we numerically implement in this paper has
been developed in Refs. \cite{Davis2001a,Davis2001b,Davis2002a,Gardiner2003a},
and here we briefly summarise the main points of this formalism, adapted
for application to inhomogeneous systems.

A dilute Bose gas is well-described by the second quantized Hamiltonian\begin{equation}
\hat{H}=\hat{H}_{{\rm sp}}+\hat{H}_{{\rm I}},\label{eq:Hfull}\end{equation}
where\begin{eqnarray}
\hat{H}_{{\rm sp}} & = & \int d\mathbf{\tilde{\mathbf{x}}\,\hat{\psi}^{\dagger}}(\tilde{\mathbf{x}})\left(-\frac{\hbar^{2}}{2m}\nabla^{2}+V_{{\rm trap}}(\tilde{\mathbf{x}})\right)\hat{\psi}(\tilde{\mathbf{x}}),\label{eq:Hsp}\\
\hat{H}_{{\rm I}} & = & \frac{1}{2}U_{0}\int d\mathbf{\tilde{\mathbf{x}}}\,\mathbf{\hat{\psi}^{\dagger}}(\mathbf{\tilde{\mathbf{x}}})\mathbf{\hat{\psi}^{\dagger}}(\tilde{\mathbf{x}})\hat{\psi}(\tilde{\mathbf{x}})\hat{\psi}(\mathbf{\tilde{\mathbf{x}}}),\label{eq:HI}\end{eqnarray}
are the single particle and interaction Hamiltonians respectively,
and $\hat{\psi}(\tilde{\mathbf{x}})$ is the quantum Bose field operator
that annihilates a particle at position $\tilde{\mathbf{x}}$. The
inclusion of an external trapping potential, assumed to be harmonic
in form i.e. \begin{equation}
V_{{\rm trap}}(\tilde{\mathbf{x}})=\frac{1}{2}m\omega_{z}^{2}(\lambda_{x}^{2}\tilde{x}^{2}+\lambda_{y}^{2}\tilde{y}^{2}+\tilde{z}^{2}),\label{eq:Vtrap}\end{equation}
is the major new feature of this paper as compared to the earlier
numerical investigations made by one of us in Refs. \cite{Davis2001a,Davis2002a}.
In Eq. (\ref{eq:HI}) particle interactions have been approximated
by a contact interaction of strength $U_{0}=4\pi\hbar^{2}a/m$, where
$m$ is the atomic mass, and $a$ is the $s$-wave scattering length.
The harmonic trap geometry is defined by the angular oscillation frequencies
along each axis, i.e. $\omega_{x}$, $\omega_{y}$, and $\omega_{z}$.
For convenience we express the $x$ and $y$ frequencies relative
to $\omega_{z}$, by introducing the relative frequency parameters
$\lambda_{x}\equiv\omega_{x}/\omega_{z}$ and $\lambda_{y}\equiv\omega_{y}/\omega_{z}$.

The exact equation of motion for the field operator is given by the
Heisenberg equation of motion\begin{equation}
i\hbar\frac{\partial\hat{\psi}}{\partial\tilde{t}}=\left(-\frac{\hbar^{2}}{2m}\nabla^{2}+V_{{\rm trap}}(\tilde{\mathbf{x}})\right)\hat{\psi}(\tilde{\mathbf{x}})+U_{0}\hat{\psi}^{\dagger}(\tilde{\mathbf{x}})\hat{\psi}(\tilde{\mathbf{x}})\hat{\psi}(\tilde{\mathbf{x}}).\label{eq:HeisenbergEOM}\end{equation}
For situations of experimental relevance the Hilbert space is enormously
large, and directly solving Eq. (\ref{eq:HeisenbergEOM}) is not possible
without further approximation. The essence of our approach is to split
the field operator into two parts representing the coherent and incoherent
regions. We define the projection operators

\begin{eqnarray}
P\{ F(\mathbf{\tilde{\mathbf{x}}})\} & = & \sum_{n\in\mathcal{C}}\varphi_{n}(\mathbf{\tilde{\mathbf{x}})}\int d^{3}\mathbf{\tilde{x}}'\varphi_{n}^{*}(\mathbf{\tilde{\mathbf{x}'})}F(\tilde{\mathbf{x}'}),\label{eq:Pop}\\
Q\{ F(\mathbf{\tilde{\mathbf{x}}})\} & = & \sum_{n\notin\mathcal{C}}\varphi_{n}(\mathbf{\tilde{\mathbf{x}})}\int d^{3}\mathbf{\tilde{x}}'\varphi_{n}^{*}(\mathbf{\tilde{\mathbf{x}'})}F(\tilde{\mathbf{x}'}),\label{eq:Qop}\end{eqnarray}
where $n\in C$ defines the modes that make up the coherent region
$C$, and $\varphi_{n}(\mathbf{\tilde{\mathbf{x}})}$ is the $n$th
eigenfunction of the basis that diagonalizes single particle Hamiltonian
$\hat{H}_{{\rm sp}}$. We define\begin{eqnarray}
\hat{\Psi}(\tilde{\mathbf{x}}) & = & P\{\hat{\psi}(\mathbf{\tilde{\mathbf{x}}})\},\nonumber \\
\hat{\eta}(\tilde{\mathbf{x}}) & = & Q\{\hat{\psi}(\mathbf{\tilde{\mathbf{x}}})\},\label{eq:fieldsplit}\end{eqnarray}
which we refer to as the coherent ($\hat{\Psi}$) and incoherent ($\hat{\eta}$)
field operators respectively. This division is based on the average
occupation of the states. The coherent field $\hat{\Psi}$ is chosen
to describe the low-lying highly occupied states of the system, \emph{i.e.}
modes containing of order five or more particles. The incoherent field
$\hat{\eta}(\mathbf{\tilde{\mathbf{x}}})$ contains the complementary
states which are sparsely occupied. Our particular interest is in
situations near equilibrium, where this separation can be conveniently
introduced via thermodynamic arguments by making an appropriate energy
cutoff $\tilde{E}_{{\rm cut}}$ in the single particle spectrum. 

To derive the PGPE, we apply the projection operator (\ref{eq:Pop})
to the equation of motion for the field operator (\ref{eq:HeisenbergEOM}).
The fundamental approximation we make, often referred to as the classical
field approximation, is to neglect the quantum mechanical nature of
coherent field operator, i.e. set $\hat{\Psi}(\mathbf{\tilde{\mathbf{x}}})\to\Psi(\mathbf{\tilde{\mathbf{x}}})$
(a c-number function) due to the high occupation numbers of these
modes. By making this approximation, the equation of motion (\ref{eq:HeisenbergEOM})
is transformed to a form involving couplings between the coherent
and incoherent fields of some complexity, (e.g. see \cite{Davis2001b}).
As a further approximation we neglect the interaction between the
coherent and incoherent regions and simply consider the equation of
motion for $\Psi(\tilde{\mathbf{x}})$ in isolation\begin{eqnarray}
i\hbar\frac{\partial\Psi(\tilde{\mathbf{x}})}{\partial t} & = & \left(-\frac{\hbar^{2}}{2m}\nabla^{2}+V_{{\rm trap}}(\tilde{\mathbf{x}})\right)\Psi(\tilde{\mathbf{x}})\nonumber \\
 &  & +\, P\bigg\{ U_{0}|\Psi(\tilde{\mathbf{x}})|^{2}\Psi(\tilde{\mathbf{x}})\bigg\},\label{eq:PGPE}\end{eqnarray}
which is the Projected Gross-Pitaevskii equation (PGPE). This equation
is of a similar form as the usual Gross-Pitaevskii equation, however
the classical field $\Psi(\tilde{\mathbf{x}})$ has a distinct interpretation:
It represents the quantum field for many low-lying modes, rather than
just the condensate mode. Several substantial approximations have
been made to reduce the full Heisenberg equation of motion for the
field (\ref{eq:HeisenbergEOM}) to the PGPE (\ref{eq:PGPE}), however
previous studies have demonstrated that this final form contains a
rich set of physics (e.g. see \cite{Davis2001a,Davis2002a,Goral2001a,Goral2002a}).

\section{Numerical Approach\label{sec:Numerical-Approach}}

The modes of the system are of central important in the assumptions
used to derive the PGPE, and care must be taken in numerical implementations
to ensure the modes are faithfully represented. It is in our opinion
that any useful simulation technique must satisfy the following requirements:

\begin{enumerate}
\item The space spanned by the modes of the simulation should match that
of the coherent region of the physical system being simulated as closely
as possible. That is, the modes should be the single particle modes
of the system up to the prescribed energy cutoff $\tilde{E}_{{\rm cut}}$. 
\item The assumption of high occupancy in all modes necessitates that the
numerical scheme must propagate all modes accurately.
\end{enumerate}
Most commonly used methods for propagating Schr\"odinger-type equations
do not satisfy these requirements, in particular many methods do not
propagate all modes of the numerical basis faithfully. This leads
to negligible errors if the highest modes are unoccupied, as is the
case for the $T=0$ GPE. However, it is clear that methods based on
such assumptions will not be appropriate for simulating the PGPE.

Before introducing the numerical scheme used in this paper, we review
how the method used by Davis \emph{et al.} \cite{Davis2001a,Davis2002a,DavisTemp}
to simulate homogeneous Bose gases addresses the aforementioned conditions.
For the homogeneous system the modes are plane-wave like and are suitable
to grid (Fourier) methods of propagation (also used in \cite{Brewczyk2004a}).
To define the coherent region Davis \emph{et al.} instigated an energy
cutoff by using an explicit projection operator in momentum space,
ensuring condition 1 was satisfied. To satisfy condition 2, \textbf{}sufficiently
many states outside the cutoff should be retained to ensure that the
nonlinear terms in the evolution equation were exactly evaluated (without
aliasing) for all modes of the coherent region. This was not realized
at the time and aliasing is present in the calculations presented
in \cite{Davis2001a,Davis2002a,DavisTemp}. However, the presence
of aliasing mainly modifies the action of the nonlinear term and does
not affect the equilibrium state or \textbf{}the conclusions reached
\textbf{}in these papers. For further details we refer the reader
to \cite{DavisAlias}.

We comment that simply using an unmodified grid method as in \cite{Brewczyk2004a}
does introduce a cutoff into the system \textbf{}in momentum space,
but does not satisfy the two criterion listed above. First, the energy
cutoff is anisotropic, varying in magnitude by a factor of three with
direction in momentum space (in three dimensions). Second, the largest
momentum states will be aliased in the calculation of nonlinear terms
and their dynamics will be misrepresented. The situation is even worse
when using an unmodified grid method to simulate classical field dynamics
for a trapped Bose gas as in \textbf{}\cite{Marshall1999a,Goral2002a,Schmidt2003a}.
It is this issue that we address here.

\subsection{Brief review of numerical method\label{sub:num_review}}

The method we have used to simulate the Projected Gross-Pitaevskii
equation (\ref{eq:PGPE}) with an explicit cutoff in energy derives
from a recent numerical scheme by Dion \emph{et al.} \cite{Dion2003a}.
We briefly review our adaption of this method.

To begin we rescale the Projected Gross-Pitaevskii equation (\ref{eq:PGPE})
by introducing units of distance $x_{0}=\sqrt{\hbar/2m\omega_{z}},$
time $t_{0}=\omega_{z}^{-1}$, and hence energy $\hbar\omega_{z}$,
with $\omega_{z}$ the trap frequency along the $z$-direction. With
these choices, we have\begin{eqnarray}
i\frac{\partial\Psi}{\partial t} & = & -\nabla^{2}\Psi+\frac{1}{4}(\lambda_{x}^{2}x^{2}+\lambda_{y}^{2}y^{2}+z^{2})\Psi+C|\Psi|^{2}\Psi,\label{eq:GPE1}\end{eqnarray}
where we have defined the nonlinear coefficient as $C=N_{C}U_{0}/\hbar\omega_{z}x_{0}^{3}$,
and for clarity have used untilded variables to indicate quantities
expressed in computational units. We take the wave function to be
normalized to unity, so that the total number of atoms within the
coherent region, $N_{C}$, appears in the definition of the nonlinearity
constant. To simplify our discussion of the numerical method, we will
take the harmonic trapping potential to be isotropic, i.e. $\lambda_{x}=\lambda_{y}=1$.
This allows us to avoid using cumbersome notation to account for different
spectral bases in each direction.

The classical field $\Psi(\mathbf{x},t)$ is expanded as \begin{equation}
\Psi(\mathbf{x},t)=\sum_{\{ l,m,n\}\in\mathcal{C}}c_{lmn}(t)\,\varphi_{l}(x)\varphi_{m}(y)\varphi_{n}(z),\label{eq:psisho}\end{equation}
where $\{\varphi_{n}(x)\}$ are the eigenstates of the 1D harmonic
oscillator Hamiltonian satisfying\begin{equation}
\left[-\frac{d^{2}}{dx^{2}}+\frac{1}{4}x^{2}\right]\varphi_{n}(x)=\epsilon_{n}\varphi_{n}(x),\label{eq:sho1D}\end{equation}
with eigenvalue $\epsilon_{n}=(n+\frac{1}{2})$. The energy cutoff
is implemented by restricting the summation indices to the set \begin{equation}
\mathcal{C}=\{ l,m,n:\epsilon_{l}+\epsilon_{m}+\epsilon_{n}\le E_{{\rm cut}}\},\label{eq:Cset}\end{equation}
with the value of $E_{{\rm cut}}$ chosen to be appropriate for the
physical system under consideration. For later convenience we define
$n_{{\rm cut}}$ to be the largest index occurring in $\mathcal{C}$,
\emph{i.e.} the quantum number of the highest energy oscillator state
in the coherent region (i.e. $n_{{\rm cut}}\approx\tilde{E}_{{\rm cut}}/\hbar\omega_{z}$).

In the basis representation, the PGPE (\ref{eq:GPE1}) takes the form\begin{eqnarray}
\frac{\partial c_{lmn}}{\partial t} & = & -i\left[\epsilon_{l}+\epsilon_{m}+\epsilon_{n}+CF_{lmn}(\Psi)\right],\label{eq:GPEshobasis}\end{eqnarray}
where\begin{equation}
F_{lmn}(\Psi)\equiv\int d^{3}\mathbf{x}\:\varphi_{l}^{*}(x)\varphi_{m}^{*}(y)\varphi_{n}^{*}(z)|\Psi(\mathbf{x},t)|^{2}\Psi(\mathbf{x},t),\label{eq:FNL}\end{equation}
is the matrix element of the nonlinear term. An important observation
made in Ref. \cite{Dion2003a} is that these matrix elements (\ref{eq:FNL})
can be computed exactly with an appropriately chosen Gauss-Hermite
quadrature. To show this we note that because the harmonic oscillator
states are of the form $\varphi_{n}(x)\sim H_{n}(x)\exp(-x^{2}/4),$
where $H_{n}(x)$ is a Hermite polynomial of degree $n$, the wave
function can be written as\begin{equation}
\Psi(\mathbf{x})=Q(x,y,z)e^{-(x^{2}+y^{2}+z^{2})/4},\label{eq:PsiQpoly}\end{equation}
where $Q(x,y,z)$ is a polynomial that, as a result of the cutoff,
is of maximum degree $n_{{\rm cut}}$ in the independent variables. 

It follows that because the interaction term (\ref{eq:FNL}) is fourth
order in the wave function, it can be written in the form\begin{equation}
F_{lmn}(\Psi)=\int d^{3}\mathbf{x}\: e^{-(x^{2}+y^{2}+z^{2})}P(x,y,z),\label{eq:FNLquad}\end{equation}
where $P(x,y,z)$ is a polynomial of maximum degree $4\, n_{{\rm cut}}$
in the independent variables. Identifying the exponential term as
the usual weight function for Gauss-Hermite quadrature, the integral
can be exactly evaluated using a three-dimensional spatial quadrature
grid of $8\,(n_{{\rm cut}})^{3}$ points. Thus we have verified that
the matrix elements $F_{lmn}(\Psi)$ can be exactly calculated. We
refer the reader to Ref. \cite{Dion2003a} for more details of how
to efficiently implement the spatial transformation and numerical
quadrature.

\section{Simulation procedure}

\subsection{Micro-canonical ergodic evolution }

The evolution of the classical field preserves several constants of
motion. These can be considered as macroscopic parameters that constrain
the micro-states available to the system. For the PGPE, the most important
such constant of motion is the total energy, given by the energy functional\begin{eqnarray}
E[\Psi] & = & \int d^{3}\tilde{\mathbf{x}}\Big[\Psi^{*}(\tilde{\mathbf{x}})\left(-\frac{\hbar^{2}}{2m}\nabla^{2}+V_{{\rm trap}}(\tilde{\mathbf{x}})\right)\Psi(\tilde{\mathbf{x}})\nonumber \\
 &  & +\frac{1}{2}U_{0}|\Psi(\tilde{\mathbf{x}})|^{4}\Big].\label{eq:Efunc}\end{eqnarray}
 Another important constant of motion is the field normalization,
given by\[
N[\Psi]=\int d^{3}\tilde{\mathbf{x}}\left|\Psi(\tilde{\mathbf{x}})\right|^{2}.\]
As discussed in section \ref{sub:num_review}, we take the classical
field to be normalized to unity for the initial condition: a scaling
choice that causes the coefficient of the nonlinear term in the PGPE
to be proportional to the initial number of particles in the coherent
region. The PGPE may have other constants of motion, such as angular
momentum components in traps with the appropriate symmetry, however
here we will only consider situations where these are approximately
zero and can be neglected. 

The lowest energy solution to the energy functional is given by the
time-independent Gross-Pitaevskii equation (e.g. see Ref. \cite{Dalfovo1999}).
We denote the energy of this solution $E_{g}$. This solution corresponds
to $T=0$, however this situation lies outside the validity regime
of the PGPE since only a single mode, i.e. the condensate mode, is
highly occupied. For application of the PGPE our interest is in regimes
with $E>E_{g}$ such that the system is at finite temperature with
many highly occupied modes.

\begin{figure}[!htb]
\includegraphics[%
  width=3.2in,
  keepaspectratio]{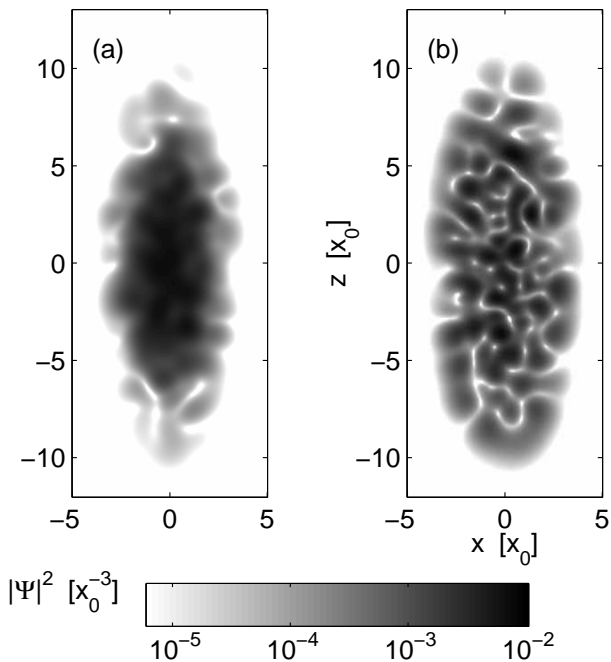}

\caption{\label{cap:evolvingfield} Thermalized classical fields for two values
of energy. Density slices taken in the $y=0$ plane for a system in
an anisotropic trap with $\lambda_{x}=\sqrt{8}$ and $\lambda_{y}=1$.
(a) Low energy case with an average energy per particle of $E=10\hbar\omega_{z}$.
(b) Higher energy case with an average energy per particle of $E=24\hbar\omega_{z}$.
For both simulations $C=2000$ and the ground state energy is $E_{g}=8.54\hbar\omega_{z}$.
The single particle cutoff energy is $E_{{\rm cut}}=31\hbar\omega_{z}$,
below which 1739 single particle states remain in the coherent region.}
\end{figure}
For the simulations we present here we make use of the ergodic hypothesis,
which we discuss further below. However, an immediate consequence
of ergodicity for the study of equilibrium properties is that the
precise details of the initial conditions in a simulation are irrelevant.
In practise we choose initial conditions to provide the desired values
for the constants of motion. The amplitudes for each single particle
mode have a random phase and occupation, but are constrained to fix
the overall normalization to unity and the energy to the desired value.
In general such an initial choice will not be a typical equilibrium
state, but under evolution the system rapidly thermalizes. This thermalization
process has been investigated for the homogeneous case in \cite{Davis2002a}.
In Fig. \ref{cap:evolvingfield} we show typical density profiles
of thermalized classical fields for the harmonically trapped system.
The cases in Fig. \ref{cap:evolvingfield}(a) and (b) differ in the
energy of the fields. It is clear from these figures that the classical
field has a somewhat chaotic appearance, and this undergoes constant
evolution as the constituent modes mix through the nonlinear interaction.

\subsection{Time-averaging correlation functions\label{sub:Time-averaging-correlation}}

The energy functional (\ref{eq:Efunc}) is nonlinear, and it is not
feasible to determine the entire \emph{phase space} of classical field
configurations consistent with a particular choice of energy. This
prohibits the calculation of quantities using an ensemble averaging
approach. However, we can determine correlation functions for the
system by using the ergodic hypothesis, i.e. replacing ensemble averages
by time-averages following the prescription\[
\langle F_{C}[\Psi(\mathbf{x})]\rangle_{{\rm ensemble}}=\lim_{\theta\to\infty}\left\{ \frac{1}{\theta}\int_{0}^{\theta}d\tau\, F_{C}[\Psi(\mathbf{x},t)]\right\} ,\]
where $F_{C}[\Psi(\mathbf{x})]$ is some functional (correlation function)
of the field. 

The general spatial correlation functions we are interested in are
usually expressed as an ensemble average over a product of quantum
field operators such as \[
F_{Q}=\langle\hat{\Psi}^{\dagger}(\mathbf{x}_{1})\ldots\hat{\Psi}^{\dagger}(\mathbf{x}_{j})\hat{\Psi}(\mathbf{x}_{j+1})\ldots\hat{\Psi}(\mathbf{x}_{n})\rangle.\]
 Note that we have only \textbf{}considered \textbf{}correlation functions
that involve the field operator for the coherent region. Also we have
restricted our attention to same-time correlations, however in principle
multi-time correlation functions could also be computed. To evaluate
these correlation functions within the framework of the classical
field theory we make the substitution $\hat{\Psi}(\mathbf{x})\to\Psi(\mathbf{x})$,
as discussed in section \ref{sec:Formalism}, and replace ensemble
averaging with time averaging. The resulting expression that we evaluate
numerically is\[
\langle F_{C}[\Psi(\mathbf{x})]\rangle_{{\rm timeave.}}=\frac{1}{N_{s}}\sum_{j=1}^{N_{s}}F_{C}[\Psi(\mathbf{x},t_{j})],\]
where $\{ t_{j}\}$ is a set of $N_{s}$ equally spaced time instances
at which the classical field has been calculated. For this choice
to well-approximate the ensemble average we require $N_{s}\gg1$,
and the time span over which averaging is done to be long compared
to the slowest time scale in the problem, e.g. the longest harmonic
oscillator period. For the equilibrium results we present in this
paper these conditions are well satisfied: we use 1400 discrete samples
of the classical field taken over an evolution time of approximately
191 oscillator periods (i.e. $t_{N_{s}}-t_{1}=1200/\omega_{z}$).

\subsection{Temperature}

For comparison with experiments and other theories it is crucial to
be able to identify the temperature of the classical field simulations
rather characterize a particular result by its energy. Previous attempts
to determine temperature have been based on fitting the occupation
of high energy modes to perturbative calculations for the spectrum
based on Hartree-Fock-Bogoliubov (HFB) theory \cite{Davis2001a,Davis2002a}.
For harmonically trapped gases, calculation of the HFB modes is much
more difficult, and limits temperature calculations to perturbative
regimes. G\'{o}ral \emph{et al.} \cite{Goral2002a} have estimated
the temperature in harmonically trapped classical field simulations
in a manner analogous to that done in experiments, by fitting the
high momentum components of the system to a noninteracting distribution.
The results of that analysis suffered from excessively large errors,
and indicate this approach would not be useful for any systematic
investigation of thermodynamic properties. \textbf{}In addition, the
high energy modes of these calculations are unlikely to have been
represented accurately \cite{Bradley}.

In recent work \cite{DavisTemp} one of us has generalized Rugh's
dynamical definition of temperature \cite{Rugh1997a} to the PGPE.
This scheme has the advantage that it is non-perturbative, and is
quite accurate. This scheme can be extended to the harmonically trapped
case and is used to calculate the temperature of the simulations presented
in this paper. Because the implementation of this scheme in the harmonically
trapped case is a trivial extension to the homogeneous implementation,
we refer the reader to Ref. \cite{DavisTemp} for details.

\section{Results}

For the results presented in this paper we have simulated a 3D system
with $\lambda_{x}=\sqrt{8},\,\lambda_{y}=1$ and an energy cutoff
of $E_{{\rm cut}}=31\hbar\omega_{z}$. This choice leads to a coherent
region containing 1739 harmonic oscillator modes.

\subsection{Condensation}

\subsubsection{Condensate fraction\label{sub:Condensate-fraction}}

Identification of the condensate fraction and mode function for an
inhomogeneous system with interactions is nontrivial. This issue was
addressed by Penrose and Onsager in 1956 \cite{Penrose1956}, who
extended the concept of Bose-Einstein condensation from an ideal gas
to the case of superfluid helium. Their primary criterion for condensation
is that a single eigenvalue, $n_{0}$, of the one-body density matrix,
$\rho(\mathbf{x},\mathbf{x}^{\prime})$, becomes an extensive parameter
of the system. With regard to obtaining a quantitative description
of the condensate, they also showed that $n_{0}$ and its corresponding
eigenvector are the condensate occupation and mode respectively.

The one-body density matrix can be written as an ensemble average
of the quantum field operators as follows:\begin{equation}
\rho(\mathbf{x},\mathbf{x}^{\prime})\equiv\langle\hat{\psi}^{\dagger}(\mathbf{x})\hat{\psi}(\mathbf{x}^{\prime})\rangle_{{\rm ensemble}}.\label{eq:rho1Brr1}\end{equation}
As we neglect the incoherent region in this paper, we restrict our
consideration to the one-body density matrix for the coherent region,
i.e. $\rho_{C}(\mathbf{x},\mathbf{x}^{\prime})\equiv\langle\hat{\Psi}^{\dagger}(\mathbf{x})\hat{\Psi}(\mathbf{x}^{\prime})\rangle_{{\rm ensemble}}$,
and using the procedure outlined in section \ref{sub:Time-averaging-correlation},
we calculate this matrix as\begin{equation}
\rho_{C}(\mathbf{x},\mathbf{x}^{\prime})\approx\frac{1}{N_{s}}\sum_{j=1}^{N_{s}}\Psi^{*}(\mathbf{x},t_{j})\Psi(\mathbf{x}^{\prime},t_{j}).\label{eq:1brr1}\end{equation}
It \textbf{}is more convenient numerically to represent the density
matrix in the spectral representation,

\begin{equation}
\rho_{ij}=\frac{1}{N_{s}}\sum_{n=1}^{N_{s}}c_{i}^{*}(t_{n})c_{j}(t_{n}),\label{eq:1bij}\end{equation}
where $c_{j}(t_{n})$ are the spectral amplitudes of the classical
field at time $t_{n}$, and the index $j$ labels all three quantum
numbers needed to specify the oscillator mode that $c_{j}$ refers
to. The efficiency of the spectral representation affords us the ability
to work with the \emph{entire} one-body density matrix. The one-body
density matrix was also time-averaged in the grid based method reported
in Ref. \cite{Goral2002a}, however their analysis was limited \textbf{}to
the $s$-wave component. 

\begin{figure}[!htb]
\includegraphics[%
  scale=0.9]{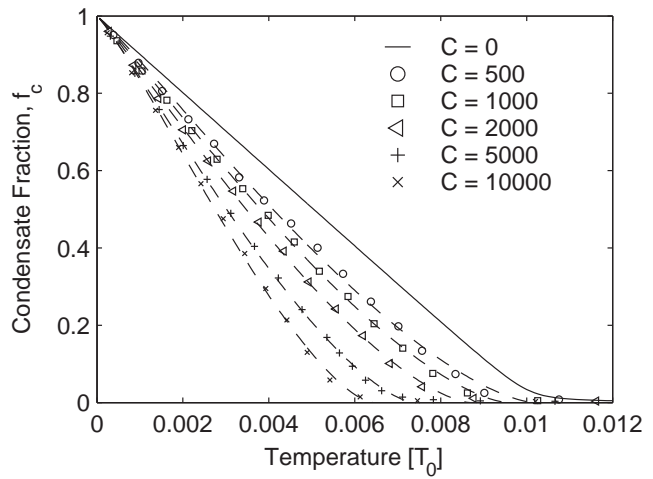}

\caption{\label{cap:n0pop} Equilibrium condensate occupation as a function
of temperature for a range of interaction strengths. The $C=0$ case
is computed using the equipartition relation \textbf{}for the modes
of the coherent region only. The \textbf{}dashed lines are simple
polynomial fits to the numerical results. Other simulation parameters
are the same as in Fig. \ref{cap:evolvingfield}. The averaging was
performed \textbf{}using $N_{s}=1400$ samples over a evolution period
of $T=1200/\omega_{z}.$ The unit of temperature is $T_{0}=N\hbar\omega_{z}/k_{B}$
\cite{DavisTemp}, where $k_{B}$ is the Boltzmann constant.}
\end{figure}
In Fig. \ref{cap:n0pop} we show the condensate fraction ($f_{c}\equiv n_{0}/N_{C}$)
as a function of system temperature for a range of interaction strengths.
The rather distinctive shape of the curves arises because these calculations
are for fixed $N_{C}$ and $E_{{\rm cut}}$. For a real system with
a fixed total number of particles, the portion of atoms in the coherent
region and the cutoff energy used to define the coherent region will
change with temperature. These issues will be important considerations
in making experimental comparisons, however \textbf{}in this paper
\textbf{}we concern ourselves with characterizing properties of the
classical field method and will only consider the case of $N_{C}$
and $E_{{\rm cut}}$ as being fixed and independent of temperature. 

For the interacting homogeneous Bose gas there is no mean field shift
of the critical temperature. Instead critical fluctuations give the
leading order shift with interaction strength to the transition temperature,
and this shift is upwards. In the trapped system there is a mean field
effect on the transition which reduces the critical temperature \cite{Giorgini1996a}.
In Fig. \ref{cap:n0pop} we observe this downward shift in the transition
temperature as the interaction strength is increased, in qualitative
agreement with recent experiments by Gerbier \emph{et al.} \cite{Gerbier2004}.
We refer the reader to \cite{DavisTempShift} for the application
of the PGPE as a quantitative model of these experiments.

\paragraph{Suitable averaging to determine the condensate fraction:}

Using linear algebra arguments it can be shown that the condensate
fraction determined according to the prescription we have outlined
above will have a lower bound of \begin{equation}
f_{c}\ge\max\{1/N_{s},1/G_{C}\},\label{eq:bound}\end{equation}
where $N_{s}$ is the number of samples used to construct the density
matrix in Eq. (\ref{eq:1bij}), and $G_{C}$ is the number of single
particle states in the coherent region. Typically we take $N_{s}<G_{C}$
so that $1/N_{s}$ forms a lower bound for the condensate fraction.
Equality in the bound holds if the individual classical fields included
in the time-average are mutually orthogonal, i.e if $\sum_{j}c_{j}^{*}(t_{m})c_{j}(t_{n})=\delta_{mn}\,\forall\,\{ m,n:\,1\le m,n\le N_{s}\}$.
Result (\ref{eq:bound}) implies, for instance, that to determine
a condensate fraction below 1\% will require us to take $N_{s}>100$
samples.

\subsubsection{Condensation: influence on density distributions in momentum and
position space}

\begin{figure*}[th]
\includegraphics[%
  width=7in,
  keepaspectratio]{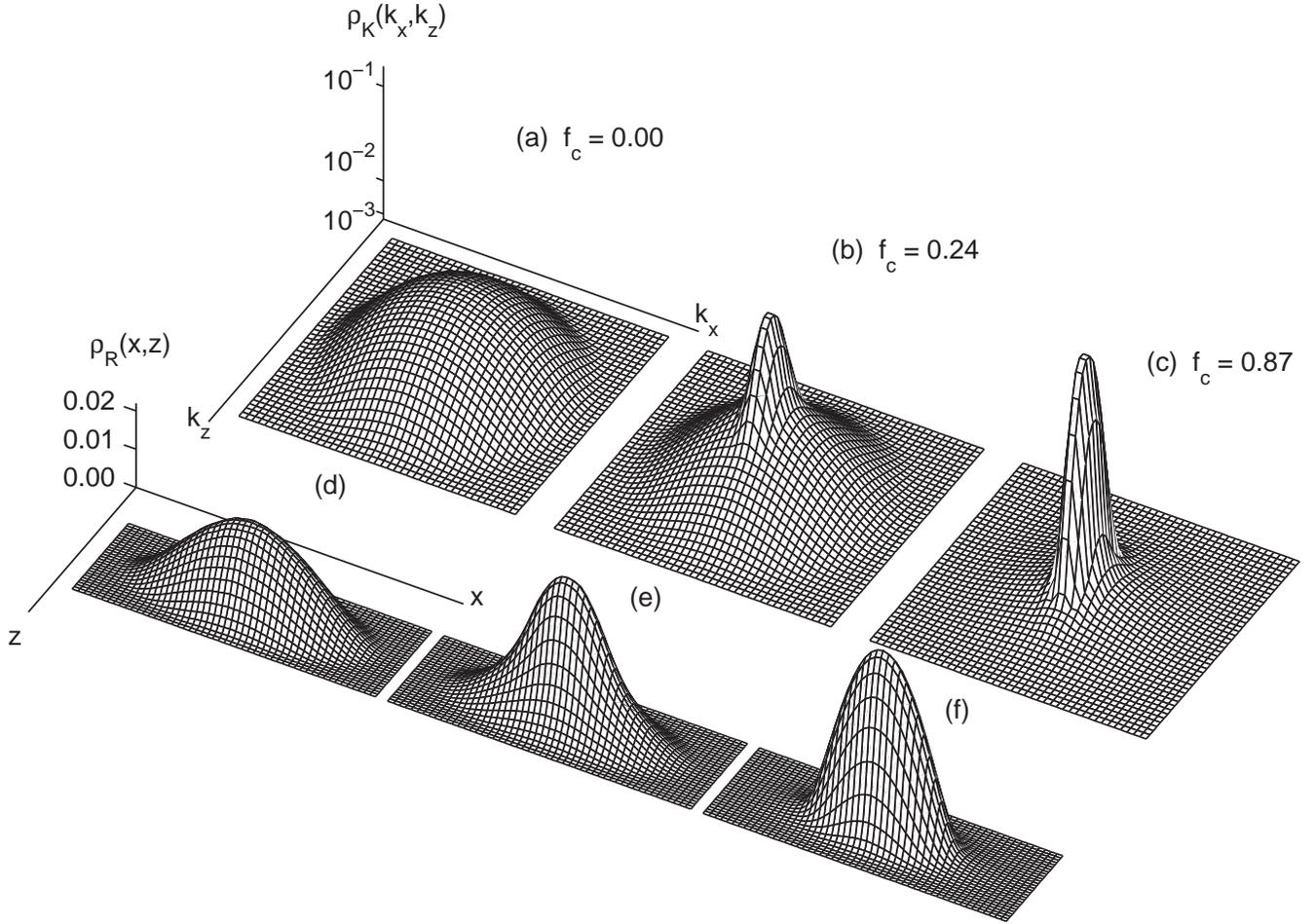}

\caption{\label{cap:mtmcondensation} Time-averaged \textbf{}column densities
in momentum (a)--(c) and position (d)--(e) space of classical field
simulations with various energies. Cases: (a) and (d) $E=24\hbar\omega_{z}$;
(b) and (e) $E=18\hbar\omega_{z}$; (c) and (f) $E=10\hbar\omega_{z}$.
Other simulation parameters are the same as in Fig. \ref{cap:evolvingfield}.
\textbf{}The averaging was performed using $N_{s}=1400$ samples over
a evolution period of $T=1200/\omega_{z}.$}
\end{figure*}

It is interesting to consider how the presence of a condensate affects
the position and momentum density profiles of the system. Indeed,
it was the appearance of an anisotropic peak in the momentum distribution
that was used as one of \textbf{}the signatures of condensation in
the first experiments \cite{Anderson1995}. In Figs. \ref{cap:mtmcondensation}(a)--(c)
we compute the momentum column density for cases above and below the
transition temperature. In detail, the quantity we calculate is the
momentum space column density given by $\int dk_{y}\,|\Phi(\mathbf{k},t)|^{2}$
(i.e. integrated along the $k_{y}$-direction), where $\Phi(\mathbf{k},t)=(2\pi)^{-3/2}\int d^{3}\mathbf{x\,}\exp(-i\mathbf{k\cdot x})\Psi(\mathbf{x},t)$
is the momentum space wave function and $\mathbf{k}=(k_{x},k_{y},k_{z})$.
Time-averaging this, i.e calculating\[
\rho_{K}(k_{x},k_{z})=\frac{1}{N_{s}}\sum_{j=1}^{N_{s}}\int dk_{y}\,|\Phi(\mathbf{k},t_{j})|^{2},\]
 yields the average column density shown in Figs. \ref{cap:mtmcondensation}(a)--(c).
The peak momentum density of the three cases considered varies over
a wide range and for presentation clarity we have used a logarithmic
density scale in Figs. \ref{cap:mtmcondensation}(a)--(c). The appearance
of a narrow peak in the momentum distribution for the condensed state
is clearly observed in Figs. \ref{cap:mtmcondensation}(b) and (c),
however the logarithmic density scale somewhat suppresses the prominence
of this feature: the peak momentum column density in Fig. \ref{cap:mtmcondensation}(c)
is 26 times larger than the peak density in Fig. \ref{cap:mtmcondensation}(a).
We also note that the momentum distribution changes from being isotropic
in Fig. \ref{cap:mtmcondensation}(a) where there is no condensate
(as calculated according to the criterion in Sec. \ref{sub:Condensate-fraction})
to exhibiting distinctive anisotropy for the condensate momentum peak
in Figs. \ref{cap:mtmcondensation}(b) and (c). This anisotropy is
directly related to the ratio of the trap frequencies. The usual experimental
method for imaging the momentum distribution is to take an absorption
image of the system after allowing it freely expand in the absence
of the trap. Interaction effects during expansion significantly suppresses
the contrast in momentum widths of condensed and uncondensed systems.
However, the contrast has been revealed by experiments using Bragg
spectroscopy techniques \cite{Stenger1999} that are able to probe
the momentum distribution \emph{in situ}. 

Similarly we can construct the column density distribution in position
space as\[
\rho_{R}(x,z)=\frac{1}{N_{s}}\sum_{j=1}^{N_{s}}\int dy\,|\Psi(\mathbf{x},t_{j})|^{2}.\]
This is equivalently obtained by integrating out the $y$-direction
of the diagonal one-body density matrix $\rho_{C}(\mathbf{x},\mathbf{x})$
(\ref{eq:1brr1}). These distributions are shown (beneath the associated
momentum distribution) in Figs.\ \ref{cap:mtmcondensation}(d)--(f).
These results emphasize that while the momentum distribution undergoes
\textbf{}substantial \textbf{}changes at the transition as discussed
above, the position distribution changes in a much more subtle manner.
The calculations presented in \cite{Goral2002a} looked only at the
behavior of slices through the position distribution of the field.

\subsection{Density Fluctuations}

Determining the condensate fraction using the Penrose-Onsager criterion
is a probe of first-order coherence in the system. Indeed, the existence
of a condensate is equivalent to \emph{off-diagonal long-range order},
i.e. the system is spatially coherent. To fully characterize the field
it is necessary to consider higher-order correlations in the system. 

To demonstrate the usefulness of the PGPE, we use it to calculate
the normalized $n$-th order coherence function at zero spatial separation,
defined as

\begin{equation}
g_{n}(\mathbf{x})=\frac{\left\langle \left(\hat{\Psi}^{\dagger}(\mathbf{x})\right)^{n}\left(\hat{\Psi}(\mathbf{x})\right)^{n}\right\rangle }{\langle\hat{\Psi}^{\dagger}(\mathbf{x})\hat{\Psi}(\mathbf{x})\rangle^{n}}.\label{eq:gn}\end{equation}
 Once again we note that we have restricted our attention to correlation
functions involving the coherent field operator. The normalized coherence
functions have been calculated for the case of $n=2$ by Dodd \emph{et
al.} \cite{Dodd1997a} using Hartree-Fock-Bogoliubov theory in the
Popov approximation, which should be valid for large condensate fractions.
In contrast the classical field result should be applicable as long
as our assumption of high mode occupancy is satisfied. 

\begin{figure}[t]
\includegraphics[%
  width=3.2in,
  keepaspectratio]{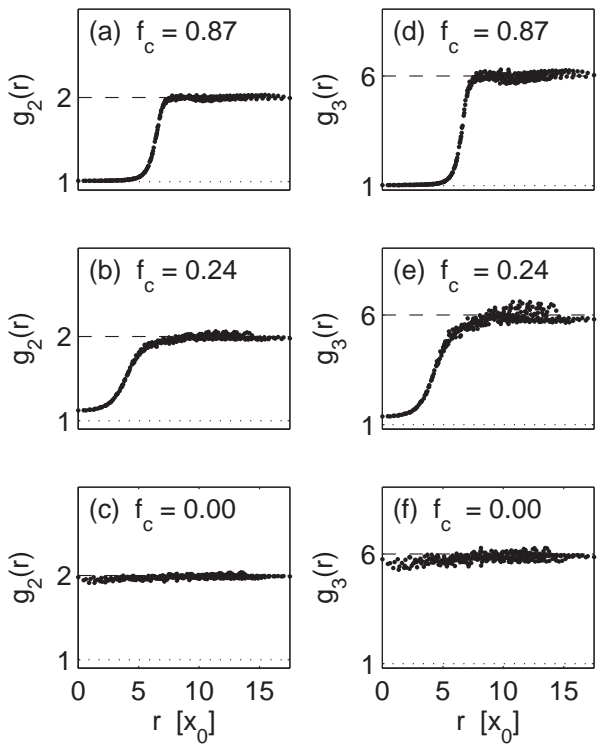}

\caption{\label{cap:spatialg2g3} Normalized coherence functions $g_{2}(r)$
and $g_{3}(r)$ (see text) as a function of radial position in the
$x=0$ plane (averaged over angle about the symmetry axis), and for
various condensate fractions. The simulation parameters are the same
as in Fig. \ref{cap:evolvingfield}, and all results are for the case
$C=2000$. For reference the coherent value ($g_{n}=1$) and thermal
value ($g_{n}=n!$) of the coherence functions are indicated as dotted
and dashed lines in the plots respectively.}
\end{figure}

Using the ergodic averaging procedure (see section \ref{sub:Time-averaging-correlation})
we evaluate (\ref{eq:gn}) for the cases $n=2$ and $n=3$. The results,
shown in Fig. \ref{cap:spatialg2g3}, are for the case of $C=2000$
in the pancake geometry trapping potential ($\lambda_{x}=\sqrt{8}$
and $\lambda_{y}=1$). For reference, we note that Fig. \ref{cap:evolvingfield}
(a) corresponds to a single profile used for the results in Figs.
\ref{cap:spatialg2g3} (a) and (d), and similarly Fig. \ref{cap:evolvingfield}
(b) corresponds to a single profile used for the results in Figs.
\ref{cap:spatialg2g3}(c) and (f). These coherence functions were
evaluated by angular averaging about the symmetry axis in the $x=0$
plane (in addition to the time-averaging), and are plotted against
the radial distance from the trap center in that plane.

The coherence functions provide a useful characterisation of quantum
fields. For instance $g_{n}=1$ for coherent fields such as a laser,
whereas $g_{n}=n!$ for thermal light fields. These features are clearly
apparent in our results for the matter-wave field. When a condensate
is \textbf{}present it occupies the center of the trapping potential
and dominates the thermal fraction of atoms in this region. This is
clearly seen in Figs. \ref{cap:spatialg2g3} (a) and (d), where for
a condensate fraction of 87\%, there is a shape transition from coherent
behavior ($g_{n}=1$) near the trap center, to thermal behavior ($g_{n}=n!$)
at a radius of approximately $r=5x_{0}$. 

In Figs. \ref{cap:spatialg2g3}(b) and (e) the same qualitative behavior
is seen, however the smaller condensate fraction (approximately 24\%
for this case) causes the boundary between coherent and thermal regions
to be closer to the trap center. Also, the coherence near the trap
center is suppressed (i.e. slightly increased from unity), indicating
that the dominant thermal fraction of the system is penetrating into
the region of the condensate, or is in some manner causing increased
fluctuations in the condensate mode. 

The results shown in Figs. \ref{cap:spatialg2g3} (c) and (f) are
for sufficiently high energy that the condensate fraction is zero.
For this case we find the thermal behavior ($g_{n}=n!$) is present
everywhere in the system.

\section{Application to non-equilibrium dynamics\label{sec:Outlook-for-application}}

\begin{figure}[tb]
\includegraphics{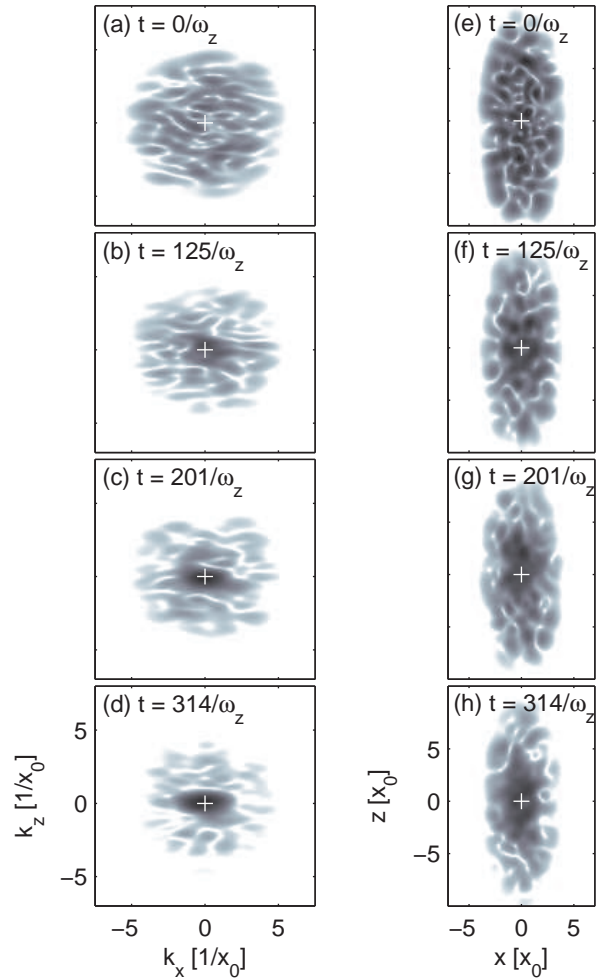}

\caption{\label{cap:Evoln} Evolution of an evaporatively cooled matter-wave
field. (a)-(d) Momentum space density in the $k_{y}=0$ plane, (e)-(h)
corresponding position space density in the $y=0$ plane. The + signs
are used to indicate thezero-coordinate in the plots for reference.
The \textbf{}initial state for the simulation is shown in Fig. \ref{cap:evolvingfield}(b),
and the evaporation is applied until $t=64\pi/\omega_{z}$ by setting
$\Psi(\mathbf{x})=0$ for $|z|>9x_{0}$ during the simulation. Other
parameters are as in Fig. \ref{cap:evolvingfield}.}
\end{figure}
 The most interesting application of the PGPE technique will be to
the non-equilibrium Bose gas. In many cases this will require a quantitative
description of the incoherent region, which is beyond the scope of
this paper, but is the subject of on going work \cite{Gardiner2003a}.
To give a qualitative demonstration of a non-equilibrium application,
we consider a simplified simulation of evaporative cooling using the
PGPE. We emphasise that this is not intended for quantitative comparison
with experiment, but the calculation highlights several interesting
features of the non-equilibrium dynamics, particularly in relation
to the identification of the condensate.

\subsection{Evaporative cooling simulation}

To perform a simulation of evaporative cooling, we begin with the
classical field in an equilibrium state above the transition temperature.
The cooling is implemented in a manner analogous to that used in experiments:
high-energy atoms that are able to venture into regions far from the
trap center are selectively removed. We do this by absorbing the portion
of the classical field which extends outside the spatial region $|z|<9x_{0}$
(i.e. setting it to zero at each time step of the simulation). We
note that for the initial state considered {[}which is the same as
the state shown in Fig. \ref{cap:evolvingfield}(b){]} only a small
fraction of the field extends into this region, and so the normalisation
and energy of the classical field is lost relatively slowly during
the cooling {[}also see Fig. \ref{cap:Evoln2}(b){]}. After 32 trap
periods the cooling mechanism is turned off and the system is allowed
to evolve and rethermalize for a subsequent 20 trap periods. A summary
of the classical field dynamics at instances during this simulation
is shown in Fig. \ref{cap:Evoln}. The initial momentum and position
profiles are shown in Figs. \ref{cap:Evoln}(a) and (e) respectively.
After approximately 20 trap periods of cooling a momentum peak has
developed in the distribution near $\mathbf{k}=\mathbf{0}$ {[}see
Fig. \ref{cap:Evoln}(b){]}. During these early stages of growth the
condensate undergoes strong sloshing and breathing dynamics as fierce
mixing occurs between the forming condensate and other low-lying quasiparticle
modes. The images in Figs. \ref{cap:Evoln}(c) and (g) show the field
at the end of the evaporative cooling ($32$ trap periods). These
figures show a large condensate centered about zero momentum {[}see
Fig. \ref{cap:Evoln}(c){]} and a relatively settled position distribution
{[}see Fig. \ref{cap:Evoln}(g){]}. The condensate exhibits breathing
dynamics, however this is significantly quenched relative to the strong
dynamics seen at earlier stages of condensate growth. The condensate
at the end of the rethermalization period is shown in Figs. \ref{cap:Evoln}(d)
and (h).

\subsection{Time-dependent condensate fraction\label{sub:-Time-dependent}}

\textgreater{}From the momentum space images of Fig. \ref{cap:Evoln}
it seems quite obvious when a Bose-Einstein condensate has formed.
A sharp peak suddenly appears in momentum space, whereas there is
no such clear signature in the real space distributions. We note that
the momentum space images are on a logarithmic scale --- so the peak
is even more obvious using a linear scale. However, in Ref. \cite{Goral2002a}
it was stated that the time-averaging inherent in the imaging process
of real experiments was essential for the splitting of the system
into a condensed and non-condensed fraction. Our results are in clear
disagreement with this conclusion, and it is at least qualitatively
apparent that condensation has occurred from a single image of the
classical field. 

\begin{figure}[t]
\includegraphics[%
  width=3in,
  keepaspectratio]{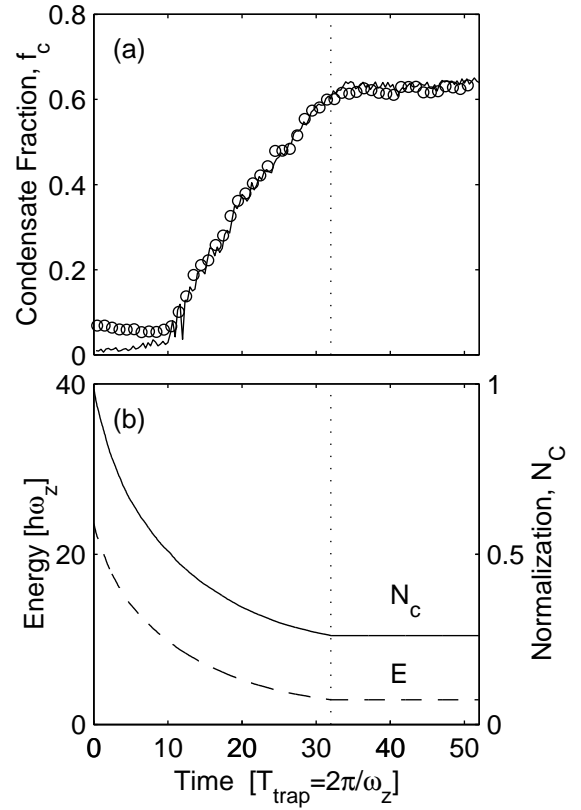}

\caption{\label{cap:Evoln2} Growth of condensate and loss of energy and normalization
during evaporative cooling process. (a) The condensate fraction of
the total remaining classical field. The open circles are calculated
by diagonalizing the one-body density matrix estimated by time-averaging
over two trap periods, while the solid curve is calculated by fitting
bimodal distributions to single-shots as described in the text. (b)
The classical field energy (dashed line) and normalization (solid
line) as a function of time. The point of time at which evaporation
is stopped is marked by a vertical dotted line. Simulation parameters
are explained in Fig. \ref{cap:Evoln}. }
\end{figure}

To quantitatively \textbf{}investigate this observation and to \textbf{}examine
the growth of the condensate, we first apply the Penrose-Onsager approach
discussed in section \ref{sub:Condensate-fraction} to the evaporative
cooling simulation. The cooling is only carried out in one dimension,
and so the dynamics proceed rather slowly. Thus it seems that we should
be able to estimate the one-body density matrix at a given time by
time-averaging over short periods. \textbf{}We calculate the condensate
fraction at trap period intervals by averaging the one-body density
matrix over that interval (by summing the classical field at 30 discrete
instances during that interval), and diagonalizing it. The results
are shown in Fig. \ref{cap:Evoln2}(a) as the open circles. Due to
the finite time over which we are able to average, the initial condensate
fraction calculated is non-zero despite the initial state having zero
condensate fraction. Because the system is not in equilibrium during
the evaporation it is not clear that the Penrose-Onsager approach
is applicable, however, the characteristic s-shaped curve we find
in Fig. \ref{cap:Evoln2}(a) is expected theoretically and has been
observed experimentally (see \cite{Kohl2002} and references therein).
Since the evaporative cooling mechanism is dissipative, particles
and energy are lost from the system. The evolution of these quantities
in the simulation are shown in Fig. \ref{cap:Evoln2}(b). This shows
that during the cooling $~74$\% of the particles and $~88$\% of
the energy in the classical field is lost.

\begin{figure}[t]
\includegraphics{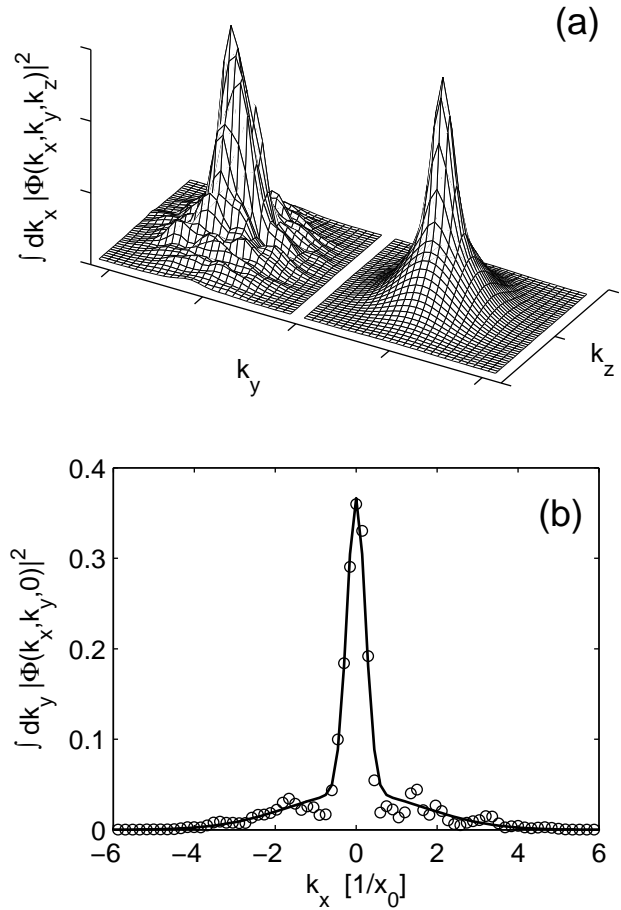}

\caption{\label{cap:fit}Example of the bimodal fitting procedure for a single-shot
of a classical field with a condensate fraction of about 0.24. (a)
The image on the left is the column density of the classical field,
while the image on the right is the bimodal fit. (b) A slice through
another column density of the same field. The open circles are the
data points and the solid line is the fitted curve.}
\end{figure}

The second approach we take is to fit bimodal distributions to single-shot
column densities of the classical field in momentum space. This computational
method mimics the actual experimental procedure that is used for fitting
condensate plus thermal cloud absorption images in the laboratory.
The fitting function is the sum of two Gaussian profiles of differing
widths, and three separate least-squares fits are carried out along
each of the $x,\textrm{}y,\textrm{ and }z$ axes. Our fitting procedure
determines the boundary of the condensate, and the condensate number
is the integral of the column density in this region less the estimated
density of the thermal cloud. It seems beneficial to fit column densities,
as this averages overs some of the fluctuations apparent in slices
through planes of the classical field, in a manner reminiscent of
the spatial averaging carried out in the realisation of a single trajectory
in \cite{Norrie2004a}. A visualisation of the results of this fitting
method are displayed in Fig. \ref{cap:fit}.

The results of this second method are also shown in Fig. \ref{cap:Evoln2}(a)
as the solid curve, and the results are in remarkable agreement with
the Penrose-Onsager approach. Thus it seems to us that the condensate
fraction in a static harmonic trap can be estimated from a single-shot
image of the classical field, without any time-averaging being necessary.
In hindsight this seems obvious, as this is the standard experimental
procedure and it seems to have had some success. The time-averaging
that occurs due to the finite exposure time in experimental imaging
is only on the order of tens of microseconds \cite{Barrettprivate},
which is almost instantaneous on the time-scale of the matter-wave
dynamics. In particular this must be the case for non-destructive
techniques such as phase-contrast imaging otherwise little information
would be gained. Intuitively the Penrose-Onsager approach should require
averaging over a time-scale of order a trap period. Thus it seems
to us that the claim that time-averaging is necessary to identify
the condensate in Ref. \cite{Goral2002a} is incorrect. Indeed, if
imaging was performed using longer duration exposures, photon recoil
effects on the atoms would dominate over any averaging of the matter
wave dynamics (e.g. see Sec. 3.5.2 of Ref. \cite{Ketterle1999a}).
A question of interest is how the methods we have used here would
perform in strongly non-equilibrium situations, however we leave this
investigation for future work.

As a final remark, we \textbf{}again \textbf{}wish to emphasise that
the simulation example neglects physical processes which would be
important in a quantitative model of an evaporative cooling experiment.
This arises through our ignorance of the incoherent region which would
be responsible for the significant transfer of particles and energy
into the coherent region. However, this model does help to illustrate
the rather complex dynamics that occur in the coherent region as it
responds to the selective removal of high energy components. The PGPE
method models the complete non-perturbative dynamics of the low-lying
modes and is naturally suited to considering non-equilibrium situations
such as evaporative cooling.

\section{Conclusions}

In this paper we have presented an efficient numerical scheme for
implementing the Projected Gross-Pitaevskii equation formalism in
three-dimensional \textbf{}harmonic traps without any axis of symmetry.
The main feature of this scheme is that it implements a consistent
energy cutoff in the harmonic oscillator basis and is suitable to
efficient and accurate numerical simulation \textbf{}on modern computer
workstations. As an application of the method we have used it to simulate
a finite temperature Bose gas in an anisotropic harmonic trap both
above and below the critical temperature. Using the ergodic hypothesis
we have obtained equilibrium quantities such as the condensate fraction
and the temperature for these simulations, and have calculated the
second- and third-order normalized coherence functions. As a non-equilibrium
application we have used the PGPE to simulate the growth of a condensate
from an evaporatively cooled thermal cloud. We have managed to identify
the condensate fraction in this calculation from both the diagonalization
of the time-averaged density matrix, as well as single-shot column
densities in momentum space of the classical field.

\section*{Acknowledgements}

PBB would like to thank Charles W. Clark of NIST for support during
the initial stages of this work, and the Otago Lasers and Applications
Research Theme for the computational resources essential to the calculations
reported. MJD acknowledges the support of the ARC Centre of Excellence
for Quantum-Atom Optics.

\bibliographystyle{apsrev}
\bibliography{projector}

\end{document}